\documentclass[prx,twocolumn,amsmath,amssymb,superscriptaddress]{revtex4-1}

\usepackage{url}
\usepackage{hyperref}
\hypersetup{
    colorlinks,
    citecolor=blue,
    filecolor=blue,
    linkcolor=blue,
    urlcolor=blue
}

\usepackage{mathrsfs}
\usepackage{verbatim}
\usepackage{graphicx}	
\usepackage{dcolumn}	
\usepackage{bm}		
\usepackage{bbm,upgreek}
\usepackage{mathtools}
\usepackage{tabularx}
\usepackage{physics}

\usepackage{xr}
\usepackage{color}
\usepackage[cal=boondoxo]{mathalfa}
\usepackage{palatino}

\usepackage{dsfont}
\usepackage{cancel}
\usepackage{xcolor}
\usepackage{harpoon}
\usepackage{accents}

\usepackage[none]{hyphenat}

\begin{document}

\title{Characterizing stochastic cell cycle dynamics in exponential growth}



\author{Dean Huang}
\author{Teresa Lo}
\affiliation{Department of Physics, University of Washington, Seattle, WA, 98195}
\author{Houra Merrikh}
\affiliation{Department of Biochemistry, Vanderbilt University, Nashville, TN, 37205}
\affiliation{Department of Pathology, Microbiology, and Immunology, Vanderbilt University Medical Center, Nashville, TN, 37232}
\author{Paul A. Wiggins}
\email{pwiggins@uw.edu}\homepage{http://mtshasta.phys.washington.edu/}
\affiliation{Department of Physics, University of Washington, Seattle, WA, 98195}
\affiliation{Department of Bioengineering, University of Washington, Seattle, WA, 98195}
\affiliation{Department of Microbiology, University of Washington, Seattle, WA, 98195}

\begin{abstract}
\noindent
Two powerful and complementary experimental approaches are commonly used to study the cell cycle and cell biology: One class of experiments characterizes the statistics (or demographics) of an unsynchronized exponentially-growing population, while the other captures cell cycle dynamics, either by time-lapse imaging of full cell cycles or in bulk experiments on synchronized populations.
In this paper, we study the  subtle relationship between observations in these two distinct experimental approaches. 
We begin with an existing model: a single-cell deterministic description of cell cycle dynamics where cell states (\textit{i.e.}~periods or phases) have precise lifetimes. We then generalize this description to a stochastic model in which the states have stochastic lifetimes, as described by arbitrary probability distribution functions. 
Our analyses of the demographics of an exponential culture reveal a simple and exact correspondence between the  deterministic and stochastic models: The corresponding state ages in the deterministic model are equal to the exponential mean of the age in the stochastic model. 
An important implication is therefore that the demographics of an exponential culture will  be well-fit by a deterministic model even if the state timing is stochastic.
Although we explore the implications of the models in the context of the \textit{Escherichia coli}  cell cycle, we expect both the models as well as the significance of the exponential-mean lifetimes to find many applications in the quantitative analysis of cell cycle dynamics in other biological systems.

\end{abstract}


\keywords{}


\maketitle
\section{Introduction}

Methods to quantitatively characterize cell cycle dynamics have expanded dramatically \cite{Willis:2017ye} since the pioneering model of the \textit{Escherichia coli} cell cycle described by Cooper and Helmstetter \cite{Cooper:1968gd}. Their initial  work represented the cell cycle as a deterministic process in which each step was precisely timed. Although these assumptions were almost certainly viewed as a matter of mathematical convenience, some later readers have interpreted the experimental success of this model 
as evidence 
that stochasticity in the cell cycle has little biological significance \cite{Bremer:1977fv}. Some later authors have relaxed some of these assumptions and found that the predictions are in fact robust to the model details \cite{Bremer:1977fv}, but none have yet reanalyzed these dynamics in the context of the significant level of stochasticity observed in cell cycle timing  (\textit{e.g.}~\cite{Wang:2010my,Robert:2014ku}). In this paper, we study a class of stochastic models that can be solved exactly, even in the strong stochasticity limit, and we explore their phenomenology. 

One fundamental difficulty with reconciling the quantitative analyses of the cell cycle is the existence of two distinct classes of experiments: In \textit{unsynchronized approaches}, an exponential culture is analyzed and the number of cells at time $t$ is used to generate statistics defined with respect to cell number \cite{Cooper:1968gd}. Examples of this approach are snapshot imaging (\textit{e.g.}~\cite*{Wang:2005ck}), flow cytometry (\textit{e.g.}~\cite{Withers:1998zt}), and many deep-sequencing based approaches (\textit{e.g.}~\cite{Rudolph:2013ho}).  We contrast these with \textit{synchronized approaches} in which cells of a known state in the cell cycle progression are analyzed.  Examples of this approach are the use of any of the previously described methods on cells which are first synchronized using a baby machine (\textit{e.g.}~\cite{Bates:2005jc}).  Time-lapse imaging of full cell cycles (\textit{e.g.}~\cite{Kuwada:2015ym}), including the use of devices like the mother machine (\textit{e.g.}~\cite{Wang:2010my}), can also be used to generate data for synchronized analyses.
Although it might  na\"ively seem that averaging with respect to these two population ensembles are equivalent, they are not. 

\begin{figure}[t]
\centering
\includegraphics[width=.9\linewidth]{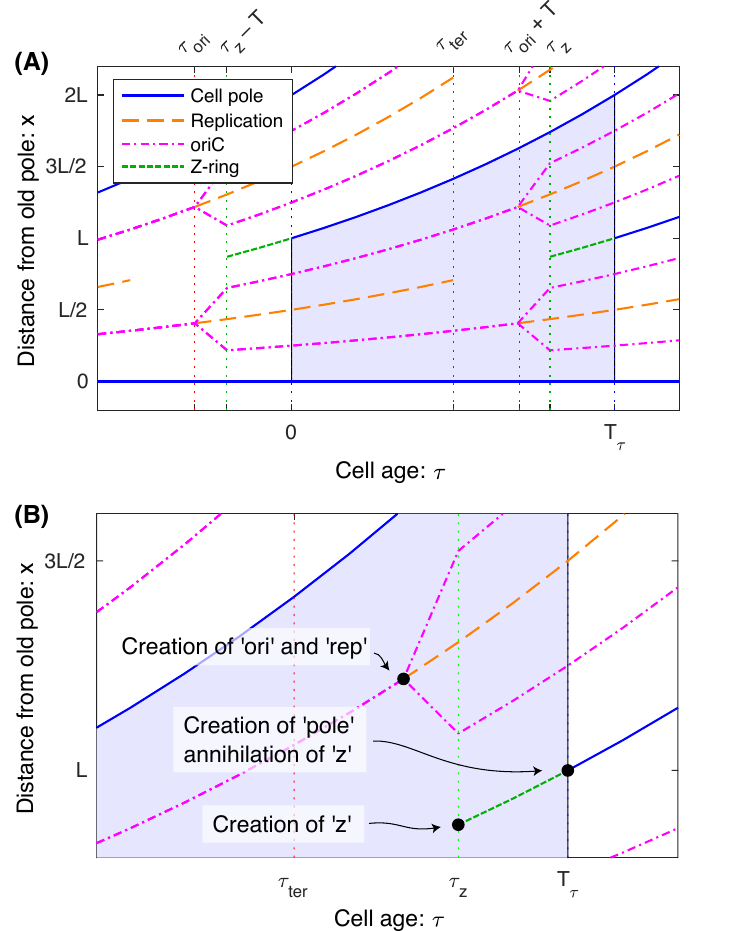}
\caption{\textbf{Panel A: Schematic for positioning and timing of events during the \textit{E.~coli} cell cycle.} In the deterministic model, the duration of the cell cycle as well as the timing of all events in the cell cycle are precise (\textit{i.e.}~deterministic).  Furthermore, we shall assume the positioning of all complexes and the length of the cell are all deterministic as well. The shaded region represents one complete cell cycle. We have also annotated the positioning and timing of a number of cell cycle events: (i) Cell poles appear as the consequence of septation and do not disappear. 
(ii) Replication of the chromosome starts when the origin (\textit{oriC}) is replicated and ends when the terminus is replicated. 
The origin is replicated before the start of the cell cycle. 
(iii) New origins are created and move from the quarter-cell positions to the eighth-cell positions after replication. (iv) The Z ring, which drives septation at midcell, assembles and disassembles at the end of the cell cycle. 
\textbf{Panel~B: Creation and annihilation.} The statistics of different types of quantities will have different statistical properties. \textit{Transient quantities} undergo both creation and annihilation events. The Z ring and replisome are natural examples of transients since they assemble and then disassemble. In contrast, both cell poles and genetic loci are \textit{perpetual quantities} that only undergo creation but never annihilation.}
\label{fig:detmodel}
\end{figure}

To demonstrate the subtlety of interpreting the data from an exponential culture, consider the probability of observing the Z ring, the  ring-shaped protein complex that forms in \textit{E.~coli} at midcell and drives the process of septation (or cytokinesis) \cite{Adams:2009cf}. If the cell cycle has  duration $T$ and the Z ring has lifetime $\delta \tau_{\rm Z}$, one might na\"ively assume the probability of observing the Z ring is:  
\begin{equation}
p_{\rm Z} = \delta \tau_z/T. \label{eqnguess}
\end{equation}
See Fig.~\ref{fig:detmodel}A. Although this is true in the synchronized population, in an exponential culture the probability is 30\% lower  as a direct consequence of the relative abundance of cells by age \footnote{The exact degree to which this is reduced depends on the ratio of $\tau_z/T$ as discussed below.}. Why? The number of new-born cells is twice the abundance of cells at the end of the cell cycle when the Z ring forms. Although this seems like a trivial book-keeping annoyance, when we consider the stochastic model, this effect has consequential implications for timing throughout the cell cycle, including on the growth rate.  

In Sec.~\ref{sec:detmodel}, we will first revisit the existing \textit{deterministic model}, where all events in the cell cycle are precisely timed. In this model, we will represent the fundamental state of the cell as an age $\tau$ and compute the statistics of cell age in an exponential culture. To make contact with observables, we then apply these results to describe the demographics of the \textit{E.~coli} cell cycle in Sec.~\ref{secappltocellcycle}.
In Sec.~\ref{secstochatic}, we consider a \textit{stochastic model}, where the cell cycle is represented as discrete sequential states $j=1...m$, each with a stochastic lifetime $\tau_{\delta j}$. Although this model cannot fully capture all the complexities of the cell cycle,  it is analytically tractable and we can exactly compute  expressions for all the same statistics as the deterministic model. The relation between the deterministic and stochastic model statistics is at this point opaque. In Sec.~\ref{eqnexpmean}, we define an \textit{exponential mean}, which is a mean biased toward younger cells that are overabundant in exponential culture. In Sec.~\ref{secmodelorr}, we demonstrate that the predictions of the stochastic and deterministic models are in fact identical if the  deterministic state ages $\tau_j$  are equal to the exponential-mean stochastic state ages $\overline{\tau}_j$.  Finally, in Sec.~\ref{secmodelorr}, we consider a number of simple biological examples to underline both the mathematical behavior of the exponential mean as well as its biological implications.  In the interest of brevity, we will discuss experimental support for this model elsewhere \cite{Lo2021b}.


%
%
%

\section{Results}
In this section, we will derive the expressions for a large number of statistics relevant for describing an exponential culture. We will first derive expressions for the statistics in the deterministic model and then the stochastic model. 
 
\subsection{Deterministic  model}

In the deterministic model, we will consider cells that are born with age $\tau = 0$ and divide deterministically at age $\tau=T_\tau$. By \textit{cell age} $\tau$, we  mean  a continuous cell state variable representing cell cycle progression, not \textit{aging} in the context of reduced cell fitness over time \cite{Stewart:2005mm}.

\label{sec:detmodel}
%

\subsubsection{Definition of the deterministic model}
In the deterministic model, cell state is described by a continuous variable, cell age $\tau$, and therefore the population is described in terms of a number density with respect to age $\tau$ at time $t$: $n_\tau(t)$.
 Age $\tau$ is defined on the interval $[0,T_\tau]$ with $\tau = 0$ corresponding to cell birth and $T_\tau$ corresponding to cell division. 
Let the cumulative creation number, $N^+_\tau(t)$, be the cumulative number of cells that have entered state (\textit{i.e.}~age) $\tau$ and the cumulative annihilation number, $N^-_\tau(t)$, be the cumulative number of cells that have transitioned out of state $\tau$ \footnote{The naming of the cumulative creation and annihilation numbers was motivated in relation to the creation and annihilation operators from quantum field theory.}. In the deterministic model, where the cell state $\tau$ is continuous (not discrete), the cumulative creation and annihilation numbers are identical and are related to the number density by the equation:
\begin{equation} \label{eq:creatnumdef}
N^+_\tau(t) = N^-_\tau(t) = \int_0^t {\rm d}t_1\ n_{\tau}(t_1).
\end{equation}
It is convenient to define these cumulative numbers in addition to the number density, since they will be a powerful tool for computing some observable quantities.
To describe the dynamics, we can write an equation describing the number of cells entering the infinitesimal age interval $[\tau,\tau+\delta \tau]$ in the infinitesimal time interval $[t,t+\delta t]$:
\begin{equation}
\delta t\, \delta \tau\, \dot{n}_\tau(t) = \delta t\, \dot{N}_{\tau}^+(t)-\delta t\, \dot{N}_{\tau+\delta \tau}^-(t),\label{eqndrate_}
\end{equation}
where $\dot{A}\equiv \partial_t A$ and the first term on the RHS is the number of cells entering state $\tau$ and the second term represents the cells leaving state $\tau+\delta \tau$. Eq.~\ref{eqndrate_} can then be rewritten:  
\begin{equation}
\dot{n}_\tau(t) = -\partial_\tau \dot{N}_{\tau}^+(t),\label{eqndrate}
\end{equation}
except at division, where some care is required.
Now consider the process of cell division explicitly: The division process can be understood as the annihilation of a cell in state $\tau=T_\tau$ and the creation of two new-born cells in state $\tau = 0$:
\begin{eqnarray}
\dot{N}_\tau^+(t) &=& \begin{cases} 2{n}_{T_\tau}(t),& \tau=0  \\
{n}_{\tau}(t),& \tau>0\label{eqn111}
\end{cases} \\
\dot{N}_\tau^-(t) &=& {n}_{\tau}(t). \label{eqn112}
\end{eqnarray}
Substituting Eq.~\ref{eqn111} into Eq.~\ref{eqndrate} gives a single piecewise rate equation in terms of the number density $n_\tau(t)$:
\begin{eqnarray}
\dot{n}_\tau(t) = -\begin{cases} 2n'_{T_\tau}(t), & \tau=0 \\
  n'_\tau (t), & \tau>0
  \end{cases},
\label{eq:ideal1} 
\label{eq:ideal2}  
\end{eqnarray}
where ${A}'\equiv \partial_\tau A$. Eq.~\ref{eq:ideal2} completely describes the cell cycle dynamics in the deterministic model.
The details of the derivation are given in Appendix~\ref{deriverate}.

\subsubsection{Solution to the deterministic model}

In steady-state growth, we can assume the total number of cells is:
\begin{equation}
N(t) = N_0 \exp( kt ),
\end{equation}
where $k$ is the growth rate that is determined by solving the rate equation (Eq.~\ref{eq:ideal2}), as detailed in Appendix~\ref{derivedet}.
It will often be convenient to rewrite the equations in terms of the  doubling time:
\begin{equation}
T\equiv k^{-1}\ln 2,
\label{defdoub}
\end{equation} 
rather than the growth rate $k$.
Eq.~\ref{eq:ideal2} evaluated at $\tau=0$ gives a consistency condition between doubling time $T$ and the duration of the cell cycle $T_\tau$:
\begin{equation}
T_\tau = T, \label{eqnidealconst}
\end{equation} 
which is to say that the doubling time is equal to the duration of the cell cycle, as one would na\"ively expect. In steady-state growth, one can compute the number density  of cells, which is:
\begin{equation}
n_\tau(t) = n_0 \exp [ k(t-\tau) ], 
\label{eq:ntaut}
\end{equation}
where $n_\tau$ is the density with respect to cell age $\tau$ and $n_0$ is a constant determined by the initial cell number. The details of the derivations for the solution and the consistency condition are given in Appendix~\ref{derivedet}.

\subsubsection{Statistics of the deterministic model}

\begin{figure}[t]
\centering
\includegraphics[width=.9\linewidth]{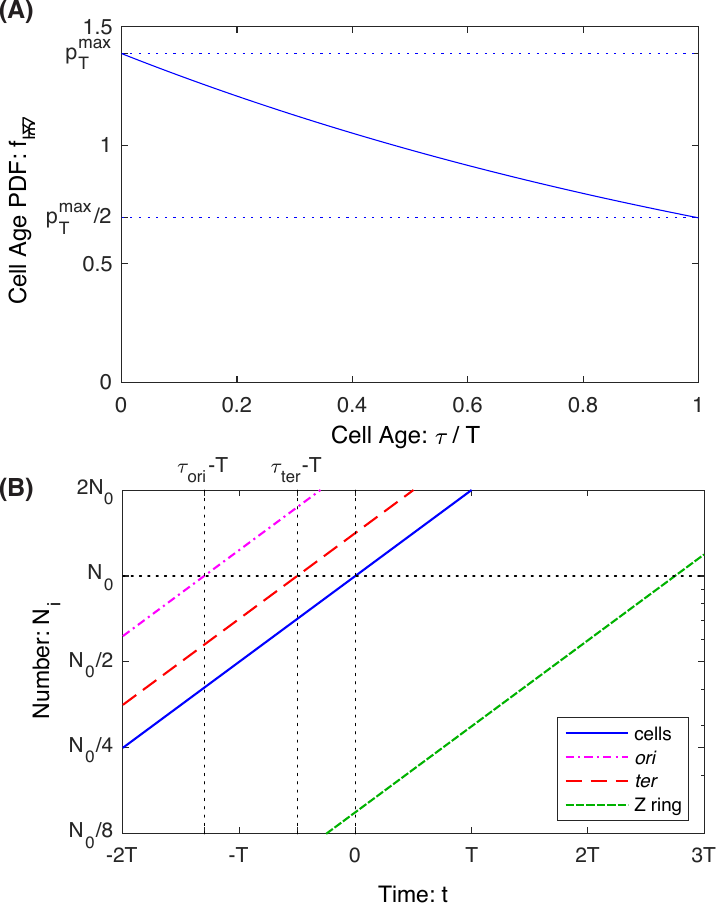}
\caption{\textbf{Panel A: Cell Age PDF.} In an exponential culture, there is an enrichment in young cells relative to old cells. The relative number of cells decays exponentially with cell age $\tau$. \textbf{Panel B: Numbers in an exponential culture.} The numbers of all quantities grow exponentially with the same growth rate. For perpetual quantities (\textit{e.g.}~cell number, cell poles, DNA loci) the relative timing of the creation of a quantity can be inferred by the temporal offset of the $N_X(t)$ curve relative to the cell number curve $N(t)$. In contrast, transient quantities, like the number of Z rings, also grow exponentially, but their offset cannot directly be interpreted as a  time. }
\label{fig:pdf}
\end{figure}

The solution of Eq.~\ref{eq:ideal2} can be used to compute the probability (PDF) and cumulative (CDF) distribution functions with respect to cell age:
\begin{eqnarray}
f_\tau(\tau) &=& 2 k e^{-k\tau}, \label{eqn:pdfage} \\
F_\tau(\tau) &=& 2 (1- e^{-k\tau}).
\end{eqnarray}
The details of the derivation are given in Appendix~\ref{derivedet_pdf}. Eq.~\ref{eqn:pdfage} implies that in an exponential culture, there is an enrichment of young cells which decays exponentially with age $\tau$. See Fig.~\ref{fig:pdf}A.

Note that the canonical observable  in an exponential culture is number as a function of time rather than abundances relative to the total number of cells $N(t)$. However, we shall write each expression as the prefactor of $N(t)$ and therefore the prefactor can be interpreted as the abundance relative to cell number $N(t)$.

The cumulative creation number is:
\begin{equation}
N^+_\tau(t) = 2e^{-k\tau} N(t). \label{eqncreation}
\end{equation}
The details of the derivation are given in Appendix~\ref{derivecreatnum}. The number of cells younger than age $\tau$ is:
\begin{equation}
N_{< \tau}(t) = 2(1-e^{-k\tau})\, N(t), \label{eqnle}
\end{equation}
and the number of cells older than age $\tau$ is:
\begin{equation}
N_{> \tau}(t) = (2e^{-k\tau}-1)\, N(t).  \label{eqng}   
\end{equation}
Finally, the number of cells in a state defined by the age range $\tau_1<\tau<\tau_2$ is: 
\begin{equation}
N_{[\tau_1,\tau_2]}(t) = N_{> \tau_1}(t)-N_{> \tau_2}(t),   \label{eqnrange}
\end{equation}
where the two terms on the right hand side are defined in Eq.~\ref{eqng}.


\subsection{Application to cell cycle dynamics}
\label{secappltocellcycle}
In this section, we demonstrate  how to apply these results in the context of the \textit{E.~coli} cell cycle dynamics shown schematically in Fig.~\ref{fig:detmodel}A. These formulae can be applied either to predict the numbers from the known replication timing or to infer timing from the observed numbers in an exponential culture.

\subsubsection{Z ring}
The Z ring is an ultra-structural complex responsible for the process of bacterial cytokinesis (or septation) in which the cell envelope contracts at midcell forming a septum that closes to form the new poles of the nascent daughter cells \cite{Du:2019qq}. The assembly, dynamics and disassembly of this structure is easily visualized using a wide range of fluorescent fusions in live cells or immunofluorescence in fixed cells \cite{Ma1997}.
 
 The Z ring is an example of a transient complex, therefore we need to use $N$ (as opposed to $N^+$). Furthermore, it assembles at $\tau = \tau_{\rm Z}$ and disassembles at the end of the cell cycle. The number of Z rings is therefore equal to the number of cells older than $\tau_{\rm Z}$:
\begin{eqnarray}
N_{\rm Z}(t) &=& N_{>\tau_Z}(t) = (2e^{-k\tau_Z}-1)\, N(t).
\end{eqnarray}
It is interesting to consider the limit as $\delta \tau_{\rm Z}\equiv T_\tau-\tau_{\rm Z}$ is small relative to the cell cycle duration $T_\tau$ in order to compare this to our intuitive guess (Eq.~\ref{eqnguess}):
\begin{eqnarray}
p_Z = \textstyle\frac{N_Z}{N} &\approx&   \textstyle\frac{\delta \tau_Z}{T}\ln 2.
\end{eqnarray}
Since $\ln 2 \approx 0.69$, this is roughly 30\% smaller than our na\"ive estimate due to depletion of older cells in an exponential culture (Fig.~\ref{fig:pdf}A).
\subsubsection{Cell poles}

Although the number of cell poles is  twice the number of cells, it is useful to consider this example more formally. Unlike the Z ring which is transient, the poles are perpetual: Once the state is created, it is never annihilated (Fig.~\ref{fig:detmodel}B). In this context, we can use the cumulative creation number $N^+$. Note that we are immediately presented with a conundrum: Are two poles formed at the end of the cell cycle ($\tau=T$) or is one pole created at birth ($\tau=0$)? Both approaches give the same number:
\begin{equation}
N_\text{pole}(t) = 2N^+_T = N^+_0 = 2N(t),
\end{equation}
which is twice the number of cells, just as one intuitively expects.

\subsubsection{DNA loci}
The numbers of DNA loci can be observed by a number of different approaches: Modern deep sequencing methods allow a replication profile (\textit{i.e.}~the DNA copy number) of all loci to be measured in a single experiment (\textit{e.g.}~\cite{Rudolph:2013ho}). However, population-level analysis of the relative copy numbers of loci long predate this modern approach \cite{Cooper:1968gd}. The single-cell dynamics of loci can also be observed: Imaging-based approaches, such as Fluorescence In Situ Hybridization and Fluorescent Repressor Operator Systems (or closely related approaches), can be used to visualize the numbers of segregated loci in single cells  \cite{Niki:2000hr,Lau:2003lc}. 

There are multiple equivalent approaches to computing the numbers of genetic locus $\ell$. First consider the slow-growth limit where both initiation and termination occur within the current cell cycle \cite{Cooper:1968gd}. Assume the locus of interest is replicated at time $\tau_\ell$. The number of copies per cell is one before replication and two after replication. We can therefore write:
\begin{eqnarray}
N_\ell(t) &=& 1\times N_{< \tau_\ell}(t)+2\times N_{> \tau_\ell}(t),\\ \label{eqnnum2l}
&=& e^{-k(\tau_\ell-T)} N(t),\label{eqnlocusrep}\\
&=& N_0 e^{k(t+T-\tau_\ell)}, \label{eqnnuml}
\end{eqnarray}
in agreement with previous results \cite{Bird:1972ux,Pritchard:1975do}. Unlike transient quantities, \textit{e.g.}~the number of Z rings, the form of Eq.~\ref{eqnnuml} implies that the number of genetic loci can  be understood as a temporal shift of $N(t)$ by $T_\tau-\tau_\ell$ to shorter times, as illustrated in Fig.~\ref{fig:pdf}B.

The cancelation between the non-exponential terms between $N_{> \tau_\ell}$ and $N_{< \tau_\ell}$ in Eq.~\ref{eqnnum2l} may seem incidental, but from another perspective it is intuitive: The mathematical reason for the non-exponential terms in the prefactors of Eqs.~\ref{eqnle}-\ref{eqng} is \textit{annihilation}, \textit{i.e.}~the reduction in the number of cells of a particular age $\tau$ due to aging. DNA loci correspond to a perpetual state: Once a locus state is created (\textit{i.e.}~replicated) it does not annihilate (\textit{i.e.}~\textit{transition} into another \textit{state}). To compute the number of genetic loci, we can therefore use the cumulative creation number $N^+$ formula (as opposed to $N$ which is reduced by annihilation). This more direct approach yields the same result as Eq.~\ref{eqnlocusrep}:
\begin{eqnarray}
N_\ell(t) &=& N^+_{\tau_\ell}(t) = e^{-k(\tau_\ell-T)} N(t), \label{eqnlocnum}
\end{eqnarray}
but is applicable for fast growth where replication initiates before the cell cycle begins (\textit{i.e.}~$\tau_\ell<0$), as illustrated in the cell cycle schematic in Fig.~\ref{fig:pdf}A.

\subsubsection{B, C, and D period}
\label{BCDperiod}

Traditionally, bacterial cell cycle is described by three periods: The B period is defined as the period between birth and replication initiation. The C period is defined as the cell cycle period during which replication occurs: \textit{i.e.} after replication initiation and before termination \cite{Cooper:1968gd}.  The D period is defined as the period between replication termination and cell division \cite{Cooper:1968gd}. There are multiple approaches to characterizing relative abundance of cells by period. A traditional approach is to infer this information from the relative abundance of the origin, terminus and number of cells \cite{Cooper:1968gd}. However, more recent single-cell approaches can visualize the replication process itself in live cells \cite{Lemon:2000rz,Wallden:2016gs}.

The relation between the durations of these periods and the locus number (Eq.~\ref{eqnlocnum}) are:
\begin{align}
B & \equiv \  \tau_\textit{ori}-0\ \ \    = T \log_2 2N/N_\textit{ori},\\
C & \equiv \  \tau_\textit{ter}-\tau_\textit{ori}  = T \log_2 N_\textit{ori}/N_\textit{ter},\\
D & \equiv \  T-\tau_\textit{ter} \ \ =  T \log_2 N_\textit{ter}/N,
\end{align}
where $N$,  $N_\textit{ori}$, and $N_\textit{ter}$ are the number of cells, origins, and termini in the exponential culture (not per cell), which has previously been  reported \cite{Cooper:1968gd,Bremer:1977fv}. Note that if replication initiates before the start of the cell cycle, $B = \tau_\textit{ori}<0$.

\subsubsection{Replication}

Finally, let us consider the replisomes and the replication process itself. A traditional population-level approach to determining the number of replicating cells is to infer it from the relative origin, terminus, and cell abundances. However, many single-cell and live single-cell approaches exist today as well. For instance, fluorescent fusions to core replisome components that localize during replication can be used to characterize the number of replicating cells \cite{Lemon:2000rz,Mangiameli:2017oc,Mangiameli:2017ex}.

Like the Z ring, replication is a transient state; however, there is a significant subtlety here: Do we count (i) replicating cells, (ii) individual replication processes consisting of replisome-pairs, or (iii) individual replisomes?

First let us consider the number of replisome-pairs. Since the replication process can span the overlap between two successive cell cycles, it is most convenient to use differences in the cumulative creation number $N^+$. In fact, we can express the number of replisome-pairs concisely in terms of  \textit{oriC} and \textit{ter}:
\begin{equation}
N_{\text{rep}}(t) = N_{\textit{ori}}-N_{\textit{ter}},
\end{equation}
and the number of individual replisomes will be twice the number of pairs. $N_{\textit{ori}}$ and $N_{\textit{ter}}$ are computed using Eq.~\ref{eqnlocnum}.

For the number of replicating cells, we consider three different cases. First consider a case where the replication cycle is internal to the cell cycle. In this case, we have:
\begin{equation}
N_{\text{rep cell}}(t) = N_{[\tau_{\textit{ori}},\tau_{\textit{ter}}]} = N_{>\tau_\textit{ori}} - N_{> \tau_\textit{ter}}
\end{equation}
which can be evaluated using Eq.~\ref{eqnrange}. If the replication process overlaps by a single cell cycle but replication rounds do not overlap:
\begin{equation}
N_{\text{rep cell}}(t) = N_{>T+\tau_\textit{ori}} + N_{< \tau_\textit{ter}}, 
\end{equation}
where $\tau_{ori}$ is negative in this context, as noted in Sec.~\ref{BCDperiod}. Finally, if the rounds of replication overlap:
\begin{equation}
N_{\text{rep cell}}(t)=N(t),
\end{equation}
and all cells are replicating in the deterministic model.

%
%
%

%
%
%
%

\subsection{Stochastic model}

\label{secstochatic}

\begin{figure}[t]
\centering
\includegraphics[width=.9\linewidth]{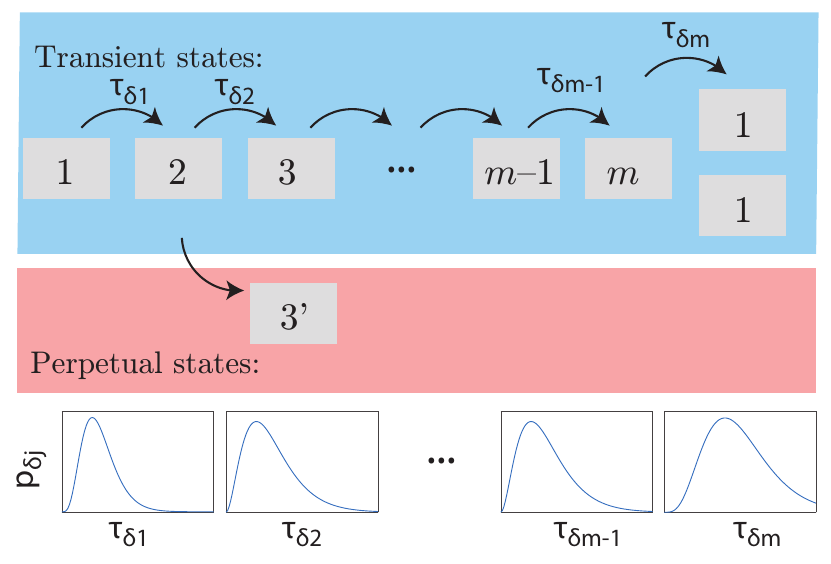}
\caption{\textbf{Schematic of the stochastic model of the cell cycle.} We represent the cell cycle as a series of sequential states $j=1...m$. After the $m$th state, two daughter cells in state 1 are born. The PDF of state lifetimes, $p_{\delta j}$, is distinct for each state $j$. Although most quantities of interest are transient, meaning that the states are \textit{created} when the  cells enter and then \textit{annihilated} when the cells exit, we also consider perpetual quantities  that are created but not annihilated (\textit{e.g.}~~DNA loci).}
\label{fig:chmemodel}
\end{figure}

An important complication of a more realistic model for cell cycle dynamics is stochasticity (\textit{i.e.}~randomness) in the lifetime of the states of the cell cycle. We will incorporate this stochasticity by dividing the cell cycle into $m$ discrete states through which the cell must transition sequentially.  This model is shown schematically in Fig.~\ref{fig:chmemodel}. 
The lifetime of each state $\tau_{\delta j}$ will be described by an arbitrary lifetime PDF, $p_{\delta j}(\cdot)$, for the $j$th state. 
It is important to note that this sequential-state stochastic model is not general enough to be an accurate representation of the bacterial cell cycle; however, it is sufficient to explore a number of interesting stochasticity-related phenomena and is exactly solvable.

\subsubsection{Definition of the stochastic model}

In our analysis, we will use the rate equation approach, rather than a master equation approach, since we are interested in the steady-state behavior of the model in the large cell number limit where the relative size of the fluctuations are vanishingly small.

Let $N_j(t)$ be the number of cells in state $j$, the cumulative creation number, $N_j^+(t)$, and the cumulative annihilation number, $N_j^-(t)$, be the total number of cells to have arrived and departed from state $j$ over all time, respectively.
The state dynamics is therefore described by the following rate equation:
\begin{equation}
\dot{N}_j = \dot{N}_j^+-\dot{N}_{j}^-.
\label{eqn:dynamics}
\end{equation} 
In this model, cells move sequentially through the $m$ states before the final state ($j=m$) transitions to the initial state ($j=1$) as two cells:
\begin{equation}
\dot{N}_{j}^+ = \begin{cases} \dot{N}_{j-1}^-, & j>1,\\
2\dot{N}_{m}^-, &j =1.
\end{cases}
\label{eqnplusminus}
\end{equation} 
Each state $j$ has a PDF of lifetimes $p_{\delta j}(t)$ and therefore the relation between state $j$ arrivals and departures is given by:
\begin{eqnarray}
\dot{N}_{j}^- &=&  p_{\delta j} \otimes \dot{N}_{j}^+,\label{conveqn}
\label{eqnconv}
\end{eqnarray}
where $\otimes$ is the convolution:
\begin{equation}
A\otimes B(t) \equiv \int^\infty_0 \!\!\! \!\!  {\rm d}t'\ A(t')\, B(t-t'). 
\end{equation}
Eqs.~\ref{eqn:dynamics}-\ref{conveqn} completely specify the stochastic model.
 
 \subsubsection{Solution to the stochastic model}

We will work in the steady-state growth limit as before. It is most convenient to work in terms of the Laplace transforms of the rate equations (Eqs.~\ref{eqn:dynamics}-\ref{conveqn}). The Laplace transform is defined:
\begin{equation}
\tilde{A}(\lambda) \equiv \int_0^\infty \!\!\!\!\! {\rm d}t\ A(t)\,e^{-\lambda t}, 
\end{equation}
where the tilde denotes the transformation from time $t$ to Laplace conjugate $\lambda$. 

The transformed representation is convenient since the ordinary differential equations become algebraic equations in terms of the transform quantities and the convolutions become products of transforms (\textit{e.g.}~\cite{Wiggins:2005gw}). Of particular importance in what follows is the relation between the PDF of lifetimes of individual state $j$, $p_{\delta j}(t)$, and the PDF of the age of the cell at the transition out of state $j$, $p_j(t)$:
\begin{equation}
\tilde p_{j} = \prod_{i=1}^{j} \tilde p_{\delta i}.
\label{eqnconvthr}
\end{equation}
A detailed derivation of the solution to the rate equations  (Eqs.~\ref{eqn:dynamics}-\ref{conveqn}) is given in Appendix~\ref{derivationstoch}.

For steady-state exponential growth, the consistency condition that relates the growth rate $k$ to the PDF of cell cycle durations $p(t)\equiv p_m(t)$ can be written concisely in terms of the Laplace transform: 
\begin{equation}
1 =   2\,\tilde{p}(k),
\label{eqn:consistgm}
\end{equation}
an equation that is well known \cite{Powell1956,PhysRevX.8.021007}. This consistency condition is equivalent to Eq.~\ref{eqnidealconst} in the deterministic model, although the mathematical equivalence between these two relations is opaque for the moment.

\subsubsection{The statistics of the stochastic model}
In an exponential culture, the cell numbers are given by the expressions:
\begin{align}
\text{Creation:} \, \ \  \ \ \  &  N^+_j(t) = 2\tilde p_{j-1}  N(t),& \label{eqnen}\\
\text{In state $\le j$:} \  \  & N_{\le j}(t) = 2[1-\tilde{p}_{j}(k)] N(t),& \\
\text{In state $>j$:} \  \  & N_{>j}(t) = [2\tilde{p}_{j}(k)-1] N(t),\label{eqnNt}& \\
\text{In state $j$:}  \, \ \ \, \  \ \        & N_j(t) = 2[\tilde{p}_{j-1}(k)-\tilde{p}_{j}(k)] N(t),& 
\end{align}
which have a similar structure to the dynamics of the deterministic model (Eqs.~\ref{eqncreation}-\ref{eqnrange}), but are dependent on the Laplace transforms of the state lifetime PDFs. Intuitively, these Laplace transforms give rise to an effective mean time.

%
%
%
%

\subsection{The exponential mean}
\label{eqnexpmean}

To understand the biological significance of the Laplace transform of the lifetime PDF, consider the generalized f-mean (or Kolmogorov mean)  where the random variable $t$ is first transformed by function $g$, an arithmetic mean is performed, and then the inverse function is applied to generate a generalized expectation \cite{Kolmogorov1930}:
\begin{equation}
\overline{t}[g] \equiv g^{-1}( \mathbb{E}_{t} g(t) ),
\label{eq:generalexpectation}
\end{equation}
where $\mathbb{E}_{t}$ is the arithmetic expectation over random variable $t$.
Both the harmonic mean and geometric mean  are special cases of this more general formulation. The Laplace transform of the lifetime and age PDFs can be reinterpreted as the expectation of $g(t) = \exp(-k t )$ and therefore we can generate the f-means:
\begin{eqnarray}
\overline{\tau}_{\delta j}(k) &\equiv& -\textstyle \frac{1}{k} \ln \tilde{p}_{\delta j}(k),\\
\overline{\tau}_j(k) &\equiv& -\textstyle \frac{1}{k} \ln \tilde{p}_{j}(k),
\label{expmeanlife}
\end{eqnarray}
which can be understood as the exponential-mean of the lifetime and age of state $j$ respectively.

Before returning to our model, we will explore the behavior of the exponential mean. 
Consider the special case of a distribution that is very narrow relative to the growth rate. In this case:
\begin{equation}
\overline{t}(k) = \mathbb{E}_t t - \textstyle{\frac{1}{2}}k \sigma^2_t + ..., \label{eqn:expan}
\end{equation}
where the exponential mean is equal to the  mean to the order of the variance ($\sigma^2_t$) times the growth rate $k$. Details of the derivation are given in Appendix~\ref{expmeannarrow}. Short-lived states and states with small lifetime-variance will therefore have exponential means equal to the mean. More generally, the Jensen inequality always guarantees the exponential mean is less than or equal to the  mean:
\begin{equation}
\overline{t}(k) \le \mathbb{E}_t t,
\end{equation}
since the function $g(t)$ is convex \cite{Jensen1906}. 

Finally, let us consider the consequences of a very wide distribution of lifetimes. Consider a state $j$ in which fraction $\epsilon$ of cells arrest ($\tau_{\delta j}\rightarrow \infty$) while the remaining cells have exponential-mean lifetime $\overline{\tau}_{\delta j,0}$. Using Eq.~\ref{expmeanlife}, it is straightforward to compute the exponential-mean lifetime:
\begin{equation}
\overline{\tau}_{\delta j} = \overline{\tau}_{\delta j,0}+T\log_2\textstyle\frac{1}{1-\epsilon}, \label{eqn:inflife}
\end{equation} 
where the second term  acts to extend the lifetime by a positive multiple of the doubling time $T$.
Although the arrested cells do lengthen the exponential-mean lifetime, it remains finite. Eq.~\ref{eqn:inflife} is a useful approximation anytime some fraction of the cells have a lifetime much longer than the doubling time even if all lifetimes are finite. 

%
%
%

\subsection{Model correspondence}

\label{modelcorrespondence}

\begin{figure}[t]
\centering
\includegraphics[width=.9\linewidth]{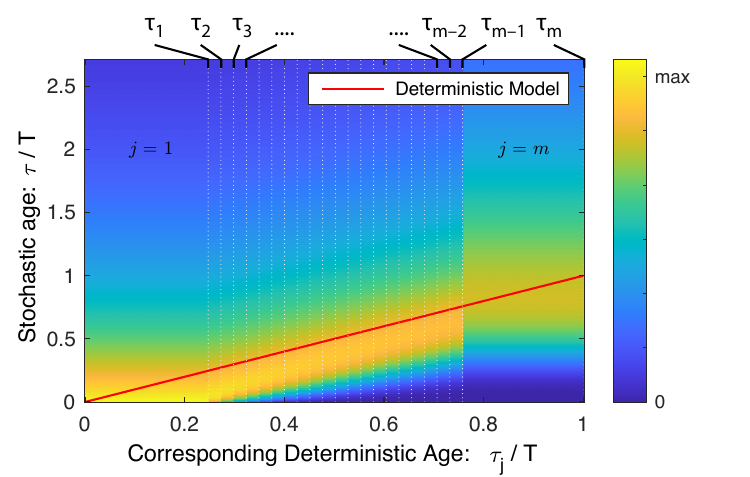}
\caption{\textbf{Model correspondence.} 
\label{fig:mrnaData} The deterministic and stochastic models generate identical statistics in exponential growth once a suitable correspondence is defined between cell state $j$ in the stochastic model and cell age $\tau$ in the deterministic model.  State $j$ in the stochastic model corresponds to age interval $(\overline{\tau}_{j-1},\overline{\tau}_{j}]$ in the deterministic model, which is represented by the red line. In the stochastic model, the PDF of age as a function of state $j$ is shown. Qualitatively, the age in the deterministic model tracks with the mode of the age in the stochastic model.}
\label{figcorr}
\end{figure}

\label{secmodelorr}
To determine the differences between the deterministic and stochastic models, we eliminate the Laplace-transformed lifetime PDFs $\tilde{p}_{\delta j}$ in favor of the exponential-mean lifetimes $\overline{\tau}_{\delta j}$ using their definition (Eq.~\ref{expmeanlife}). First consider the consistency condition for exponential growth (Eq.~\ref{eqn:consistgm}).  The convolution theorem  ensures that the exponential-mean lifetimes of successive states add to generate the age of state $j$ (\textit{e.g.}~the natural logarithm of Eq.~\ref{eqnconvthr}). 
 Eq.~\ref{eqn:consistgm} can now be rewritten as a relation between the exponential-mean cell-cycle duration and the doubling time:
 \begin{equation}
 \overline{T}_\tau \equiv \sum_{i=1}^m\overline{\tau}_{\delta i} = T,\label{eqsumoft}
 \end{equation}
 which is now intuitively equivalent to the consistency condition in the deterministic model (Eq.~\ref{eqnidealconst}). Details of the derivation are given in Appendix~\ref{stoch_consistcond}.%
 
Now consider the expressions for state number  in the stochastic model (Eqs.~\ref{eqnen}-\ref{eqnNt}).
 When the deterministic age $\tau$ is evaluated at exponential-mean stochastic age  $\overline{\tau}_j$, the  numbers are identical in the two models:
\begin{align}
\text{Entered $j$:} \, \ \  \ \ \  &  \left.N^+_j(t) = N^+_{\tau}(t)\right|_{\tau=\overline{\tau}_{j-1}} ,& \\
\text{In state $\le j$:} \  \  & N_{\le j}(t) = \left.N_{\le \tau}(t)\right|_{\tau=\overline{\tau}_{j}} ,& \\
\text{In state $> j$:} \  \  & N_{>j}(t) = \left.N_{> \tau}(t)\right|_{\tau=\overline{\tau}_{j}},& \\
\text{In state $j$:}  \, \ \ \, \  \ \        & N_j(t) = N_{[\tau_{j-1},\tau_{j}]}(t).  \label{eqnapprox}
\end{align}
%
We therefore conclude that the statistics of the deterministic and stochastic models are identical in an exponential culture for models with the same growth rate $k$, once a correspondence has been established between states $j$ and ages $\tau$. In the deterministic model, state $j$ corresponds to times $\tau \in [\overline{\tau}_{j-1},\overline{\tau}_{j}]$ where $\overline{\tau}_{0}\equiv 0$ and $\overline{\tau}_{m}=T$. This correspondence is illustrated schematically in Fig.~\ref{figcorr}.
Since we demonstrated a correspondence between the models, almost all the application discussed in Sec.~\ref{secappltocellcycle} generalize by replacing the deterministic time $\tau$ with the corresponding exponential mean $\overline{\tau}$ \footnote{ The exception are the replicating cell statistics $N_\text{rep cell}$. In this case, the implicit assumption that states are sequential cannot always be implemented in the stochastic model. For instance, consider the re-initiation of replication at the quarter cell positions late in the cell cycle in rapidly proliferating cells, as illustrated in  Fig.~\ref{fig:detmodel}A. If these events were modeled as independent, you will first see one replisome initiate at one quarter cell position and then the other. In this case one must compute the exponential means using the order statistics (\textit{e.g.}~\cite{Cox1974}).}.

%
 
 \label{sec:cdf}

%

\begin{table}
\begin{tabularx}{.48 \textwidth}{X|X|X}
Strain & Doubling Time: $T$ & C period: $C$ \\
\hline
\hline
Wildtype & $T_{\rm W}$ & $C_{\rm W}$ \\
Mutant A &  $T_{\rm W}+\frac{\epsilon}{\ln 2}T_{\rm W}$ & $C_{\rm W}+ \frac{\epsilon}{\ln 2}T_{\rm W}$ \\
Mutant S &  $T_{\rm W}+\epsilon'\, C_{\rm W}$ & $C_{\rm W}+ \epsilon'\, C_{\rm W}$ \\
\end{tabularx}
\caption{ The effect of mutants on the doubling time $T$ and C period duration $C$ of an exponential culture. \label{Tablength}}
\end{table}

\subsection{Implications for cell cycle phenomenology}
\label{sec:last}

To explore the nontrivial consequences of  stochasticity in timing, consider an example motivated by replication conflicts \cite{Merrikh:2012ez}: By visualizing the replisome dynamics using single-molecule microscopy, we have recently reported that transcription leads to pervasive replisome instability \cite{Mangiameli:2017oc}. To what extent should conflict-induced pauses in replication have been detectable in the classic analyses of  unsynchronized cell populations? 

 Consider a simplified model in which an experiment probes the difference between the wildtype strain W and two mutant strains. The wildtype W grows with deterministic C period $C_{\rm W}$ and deterministic cell cycle duration $T_{\rm W}$. In mutant strain A(rrest), a small fraction $\epsilon$ of cells arrest during replication (\textit{i.e.}~C period) and never complete the cell cycle, whereas non-arrested cells are identical to wildtype cells. In mutant strain S(low), the replication process is $1+\epsilon'$ times slower,
but the B and D periods are identical to wildtype. Using Eq.~ \ref{eqn:inflife}, one can compute the C period duration $C$ and doubling time $T$. To lowest order in $\epsilon$, the inferred cell cycle durations and C period are presented in Tab.~\ref{Tablength}.

At an intuitive level, one aspect of the prediction is easy to understand: In both mutants the C period is lengthened, as one might na\"ively expect since this is the replication period of the cell cycle. Furthermore, the doubling time increases by the same amount as the C period increases. But there is an aspect of this prediction which is perhaps less intuitive: One might na\"ively expect to observe a more dramatic consequence of replication arrest, like a large buildup of C period cells, but the consequences are indistinguishable from a slowdown in an exponential culture. In both mutants, there is a slight lengthening of the inferred C period, even though the slowdown is caused by replication arrest in the context of the A mutant. Although this prediction is not new in a qualitative sense, it concisely illustrates how the statistics of the exponential culture mask two mechanistically distinct phenomena.

The statistics of an exponential culture can also generate distinctions where seemingly none exist. Consider a more realistic model in which the duration of the D period is stochastic, has a non-zero width, and is \textit{identical} for all three strains. The more rapid growth rate of the wildtype strain implies that its effective D period is shorter than for mutants cells:
\begin{equation}
D_{\rm S},D_{\rm A} > D_{\rm W},
\end{equation}
even though the distributions of the D period durations are identical in all three strains. (To understand how this occurs, see the second term on the RHS of Eq.~\ref{eqn:expan}.) 
In most cases this effect should be subtle, but for large changes in growth rate, these changes could be quite significant and  can clearly complicate the interpretation of effective period lifetimes in an exponential culture.


\section{Discussion}

In this paper we provide a detailed analysis of  both  deterministic and stochastic models of the cell cycle. In Sec.~\ref{sec:detmodel}, we solved the deterministic model in which the cell-aging and division processes are precisely timed and determined the demographics (\textit{i.e.}~statistics) of an exponential culture. Given a set of observed demographics, Sec.~\ref{secappltocellcycle} provides a detailed road map for how to infer cell cycle state timing in the context of the deterministic model. In Sec.~\ref{secstochatic}, we solved the more realistic stochastic model in which the lifetimes of sequential states are stochastic and again we determined the demographics of an exponential culture. By defining an exponential mean in Sec.~\ref{eqnexpmean}, we demonstrated that the statistics of the two models were equivalent in Sec.~\ref{modelcorrespondence}. The effective lifetime of states in the deterministic model is the exponential mean of the lifetimes in the stochastic model. That is to say that the exponential-mean lifetimes are the \textit{sufficient statistics} of the model (\textit{e.g.}~\cite{Cox1974}): Knowledge of only these lifetime statistics predicts  the demographics of the exponential culture; therefore, inference on  exponential-culture demographics infers only the exponential means, rather than the underlying lifetime distributions themselves. Finally in Sec.~\ref{sec:last}, we discussed some of the limitations of the exponential-mean lifetimes in resolving the underlying biological mechanisms.

\subsection{Applicability of the stochastic model}
Is the stochastic model sufficiently complex to capture  all the relevant cell cycle phenomenology in \textit{E.~coli} and other bacterial systems? Like the deterministic model before it, the stochastic model is an idealized model that is simple enough to be tractable analytically, but complex enough to capture some important phenomenology.
There are a number of shortcomings of this model but perhaps the most significant is that there is \textit{no memory} beyond the cell state index $j$. As a consequence, it makes predictions at variance with some observed phenomenology: For instance, 
the stochastic model must predict that successive cell cycle durations are uncorrelated; however, these correlations are observed \cite{Wang:2010my,lin_amir_2017}. (We briefly consider the implications of a more general model in Appendix \ref{genmodelsec}.)
 Another important limitation of the stochastic model is that cell divisions  are symmetric, which is a good approximation in \textit{E.~coli}, but these types of stochastic models can easily be extended to the general asymmetric division case (\textit{e.g.}~\cite{PhysRevX.8.021007}).


\subsection{On the applicability of the exponential mean}

Although the definition of the exponential mean was motivated by the correspondence between the deterministic and stochastic models, it almost certainly has much greater applicability to other more complicated scenarios. For instance, our own numerical experiments using more complex models suggest that the relation between the effective lifetime of the states and the exponential-mean lifetime appears to be more robust than the assumptions of the stochastic model might imply. 
Since the key mechanism for generating bias toward short times is steady-state exponential growth, we expect the exponential mean of wait times to be the determinative statistic in more general models, as demonstrated in Appendix \ref{genmodelsec}. As such, the exponential-mean lifetime could be a powerful observable to  bridge timescales between single-cell and culture phenomenology in two different contexts: (i) in experiments probing cell cycle dynamics at the single-cell level and (ii) in complex numerical simulations that are too slow and too memory intensive to simulate in the long time limit \footnote{ Personal communication from S. Iyer-Biswas.}.

We should note that although we believe our interpretation of the doubling time as an exponential mean (Eq.~\ref{eqsumoft}) is novel, it has already been appreciated in two important respects: (i) From a computational perspective, the Laplace-transform formulation (Eq.~\ref{eqn:consistgm}) of Eq.~\ref{eqsumoft} has long been known \cite{Powell1956}. (ii) From a qualitative perspective, biologists have long understood the consequences of the exponential-mean lifetime on cell growth rate: \textit{I.e.} the  doubling time $T$ is ``an average'' of the cell cycle duration $T_\tau$; however, a small arrested subpopulation, for whom $T_\tau\rightarrow \infty$, slows but does not stop growth. 
 There is also physical precedent for this type of mean: Intriguingly, it emerges in context of non-equilibrium statistical mechanics \cite{Jarzynski1997} \footnote{ Personal communication from S. Iyer-Biswas.}, although what connection this has to our cellular dynamics is opaque.

\subsection{On the significance of stochasticity}

How does stochasticity affect biological function? Experimentally, we have long known that  although  the statistics of an exponentially growing population are well described by the deterministic model \cite{Cooper:1968gd}, nontrivial stochasticity in cell cycle timing is observed \cite{Wang:2010my,Robert:2014ku}. It is therefore tempting to conclude, based on the literature and perhaps even our own results,  that  stochasticity is either \textit{small} or simply \textit{does not} significantly affect biological function.

Our own conclusions are much more nuanced. Although our results guarantee that the deterministic model fits the exponential-culture demographic data just as well as the stochastic model, we have demonstrated  that the stochasticity in timing is hidden in plain sight.  The distribution of state lifetimes  determine the exponential means.
Therefore, the success of the deterministic model should not be interpreted as evidence against stochasticity or against its importance, but rather it indicates that only the exponential-mean state lifetimes are determinative parameters in the model for the demographics of an exponential culture. 

Perhaps more than anything else, the exact correspondence between the deterministic and stochastic models emphasizes the need for synchronized single-cell measurements: In Sec.~\ref{sec:last}, we illustrated  (i) how similarities in the effective duration of the C period obscures distinct biological mechanisms as well as (ii) how differences in the effective D period could belie an identical mechanism. 

At a mechanistic level, stochasticity  plays a central role in many processes. For instance, the mechanism that restarts replication will prevent the existence of a \textit{fat tail} on the distribution of C periods \cite{Mangiameli:2017oc,Merrikh:2011cq,Merrikh:2012ez}. Although the existence of the fat tail---\textit{i.e.}~a small number of cells with very long C periods---does not \textit{break} the correspondence with the deterministic model, it does increase the exponential-mean C period, which in turn decreases the growth rate. (E.g. see Tab.~\ref{Tablength}.) Since the growth rate is decreased, there is a strong selective pressure to reduce  \textit{stochasticity}. This argument predicts the existence of biological mechanisms to reduce stochasticity, as are already known in many contexts (\textit{e.g.}~replication restart). 
In fact, the subtle signature of stochasticity  suggests an interesting hypothesis: a significant number of mutants that are currently known to reduce growth rate may in fact generate this phenotype by increasing the level of stochasticity in the cell cycle duration. Single-cell experimental analysis must play a central role in understanding these phenomena. 

\medskip

\acknowledgements

PAW acknowledges advice and comments from M.~Cosentino-Lagomarsino, S.~Iyer-Biswas, P.~Levine, J.~Mittler, R.~Phillips, M.~Transtrum, B.~Traxler, I.M.~Shelby, and H.K.~Choi. This work was supported by NIH grant R01-GM128191.

%

\appendix

\section{Supplemental derivations}

\subsection{Derivation of the rate equation in the deterministic model}

\label{deriverate}
To obtain the cumulative creation and annihilation numbers in terms of the number density (Eq.~\ref{eq:creatnumdef}), we integrate the number density at fixed age $\tau$ over all time $t$ to obtain the cumulative number of cells that have ever entered (creation) or left (annihilation) age $\tau$. They are equivalent due to the continuous nature of the deterministic model. If states were discrete, as in the stochastic model, then the cumulative creation and annihilation numbers would differ by the number of cells currently in the state $\tau$. 

To obtain Eq.~\ref{eqndrate} from Eq.~\ref{eqndrate_}, we divide both sides of Eq.~\ref{eqndrate_} by $\delta t\,\delta\tau$ and replace $\dot{N}_{\tau+\delta\tau}^-(t)$ with the equivalent $\dot{N}_{\tau+\delta\tau}^+(t)$, which leaves:
\begin{equation}
    \dot{n}_\tau(t)=-\textstyle\frac{\dot{N}_{\tau+\delta\tau}^+(t)-\dot{N}_\tau^+(t)}{\delta\tau}.
\end{equation}
Taking the limit as $\delta\tau$ goes to 0 and using the definition of a derivative, we are left with:
\begin{equation}
    \dot{n}_\tau(t)=-\partial_\tau\dot{N}_\tau^+(t),
\end{equation}
which is Eq.~\ref{eqndrate} in the main text. Eqs.~\ref{eqn111}-\ref{eqn112} follow from taking the partial time derivative of Eq.~\ref{eq:creatnumdef} and taking into account the consistency condition:
\begin{equation}
    n_0(t)=2n_{T_\tau}(t).
\label{eq:consistcond}
\end{equation}
Conceptually, this consistency condition describes how cell division at age $T_\tau$ leads to twice as many daughter cells of age ${\tau=0}$.

\subsection{Derivation of the solution in the deterministic model}

\label{derivedet}
 In the deterministic model, we can assume that steady-state growth of the population is represented by an exponentially increasing time dependence factor, $e^{kt}$, with a constant unknown growth rate $k$. This assumption holds in the long time limit, since only the fastest growing mode remains in exponential growth, while all others (smaller $k$) are diluted out. We thus stipulate that in the deterministic model, all cellular quantities must grow with this same time dependence. The number density is then a solution of the form: 
\begin{eqnarray}
n_\tau(t) = n_\tau(0)\, e^{kt},
\end{eqnarray}
where $n_\tau(0)$ represents the initial age distribution at ${t=0}$. Plugging this into Eq.~\ref{eq:ideal1} for the ${\tau>0}$ case yields:
\begin{eqnarray}
    k\,n_\tau(t)=-\textstyle\pdv{}{\tau}\qty(n_\tau(t)).
\end{eqnarray}
This can then be integrated to yield the solution:
\begin{eqnarray}
n_\tau(t) &=& n_0(t)\, e^{-k\tau},\\
&=& n_0\,e^{kt}e^{-k\tau},
\label{eq:ideal3} 
\end{eqnarray}
where $n_0$ is a constant determined by the initial cell number. This equation appears in the main text as Eq.~\ref{eq:ntaut}.
To satisfy the $\tau=0$ case of Eq.~\ref{eq:ideal1}, we must use the consistency condition (Eq.~\ref{eq:consistcond}):
\begin{eqnarray}
n_0(t)&=&2n_{T_\tau}(t),\\
n_0e^{kt}&=&2n_0e^{kt}e^{-kT_\tau}.
\end{eqnarray}
Dividing both sides by $n_0e^{kt}$ and solving for $T_\tau$ gives:
\begin{eqnarray}
T_\tau&=&k^{-1}\ln2.
\end{eqnarray}
Therefore, the doubling time defined in Eq.~\ref{defdoub} is equivalent to the cell cycle duration, as one would na\"ively expect. Furthermore, this equation relates the growth rate $k$, a population measure, to the cell cycle duration $T_\tau$, a single-cell measure.

\subsection{Derivation of the PDF and CDF in the deterministic model}

\label{derivedet_pdf}
To obtain the probability distribution function, $f_\tau(\tau)$, with respect to cell age (Eq.~\ref{eqn:pdfage}), we must normalize the number density at any fixed time $t$:
\begin{eqnarray}
f_\tau(\tau)=\frac{n(\tau)}{\int_0^{T_\tau}n(\tau)\dd \tau},
\end{eqnarray}
where $n(\tau)=n(\tau=0)\,e^{-k\tau}$, which is just Eq.~\ref{eq:ntaut} with the fixed $t$ factor absorbed into ${n(\tau=0)}$. Evaluating the integral and replacing $n(\tau)$ with the expanded form yields:
\begin{eqnarray}
f_\tau(\tau)&=&\frac{n(0)\,e^{-k\tau}}{-n(0)\frac{1}{k}(e^{-kT_\tau}-1)},\\
&=&\frac{ke^{-k\tau}}{1-e^{-kT_\tau}}.
\label{eq:pdfderive}
\end{eqnarray}
Now recall that ${T_\tau=T\equiv k^{-1}\ln2}$, from Eqs.~\ref{defdoub}-\ref{eqnidealconst}. Plugging this $T_\tau$ into Eq.~\ref{eq:pdfderive} gives Eq.~\ref{eqn:pdfage}:
\begin{equation}
f_\tau(\tau) = 2 k e^{-k\tau}.
\end{equation}
To obtain the CDF, integrate with respect to $\tau$ from $0$ to $\tau$:
\begin{eqnarray}
F_\tau(\tau) &=& \textstyle\int_0^\tau 2ke^{-k\tau'}\dd\tau'\\
&=& 2 (1- e^{-k\tau}).
\end{eqnarray}

\subsection{Derivation of the cumulative creation number in terms of $N(t)$}

\label{derivecreatnum}
Consider the cumulative creation number Eq.~\ref{eq:creatnumdef} evaluated at $\tau=0$:
\begin{equation}
N^+_0(t) = 2 N(t), \label{eqncreation2}
\end{equation}
which is double the current number of cells $N$.
The factor of two arises due to the cumulative nature of $N_\tau^+(t)$. To understand this intuitively, consider the total number of cells in each generation:
\begin{equation}
N^+_0(t) = (1+\textstyle \frac{1}{2}+ \frac{1}{4} + ... )N(t), \label{eqncreation3}
\end{equation}
which is a geometric series and can be summed to $2N$, matching Eq.~\ref{eqncreation2}. We can now multiply by the $\tau$-dependence term, $e^{-k\tau}$, to obtain:
\begin{equation}
N^+_\tau(t) = 2e^{-k\tau} N(t). \label{eq:appendixcreatnum}
\end{equation}
More formally, we can use direct integration of Eq.~\ref{eq:ntaut}:
\begin{eqnarray}
N_\tau^+(t) &=& \textstyle\int n_\tau(t')\dd t',\\
&=& \textstyle\frac{1}{k}n_0e^{-k\tau}e^{kt}+c,\label{eq:creatnumint2}\\
N_0^+(t) &=& \textstyle\frac{1}{k}n_0e^{kt}+c,\label{eq:creatnumint3}
\end{eqnarray}
where c is an integration constant. We now integrate $n_\tau(t)$ over all $\tau$ to obtain the total number of cells at time $t$:
\begin{eqnarray}
N(t) &=& \textstyle\int_0^{T_\tau}n_{\tau'}(t)\dd \tau',\\
&=& \textstyle-\frac{1}{k}n_0e^{kt}\qty(e^{-kT_\tau}-1),\\
&=& \textstyle\frac{1}{2k}n_0e^{kt}.
\end{eqnarray}
Combining this with the consistency condition Eq.~\ref{eqncreation2}, we get:
\begin{equation}
N_0^+(t)=\textstyle\frac{1}{k}n_0e^{kt}.
\end{equation}
Setting this equal to Eq.~\ref{eq:creatnumint3} allows us to set the integration constant ${c=0}$. Eq.~\ref{eq:creatnumint2} then becomes Eq.~\ref{eq:appendixcreatnum} from above, which is Eq.~\ref{eqncreation} from the main text.





\subsection{Derivation of the solution in the stochastic model}

\label{derivationstoch}

Clearly, Eqs.~\ref{eqnplusminus} and \ref{eqnconv} can be combined recursively to generate a relation between the number of cells entering states 1 and $j$. First let us define the state-transition time PDF $p_j$, describing the total time taken to transition from birth through state $j$:
\begin{eqnarray}
\tilde p_{j} &\equiv& \prod_{i=1}^{j} \tilde p_{\delta i},\\
\tilde{p} &\equiv&  \tilde p_{m} 
\end{eqnarray}
where $p$ is the lifetime PDF for the entire cell cycle.
We then write an expression of the number arriving in state $j$:
\begin{equation}
\tilde{N}_{j}^+ =   \tilde p_{j-1} \tilde{N}_{1}^+.\label{eq1plus}
\end{equation}
As before, using this same condition at the end of the cell cycle gives rise to a consistency condition:
\begin{equation}
\tilde{N}_{1}^+ =   2 \tilde{p}  \tilde{N}_{1}^+.
\end{equation}
It follows that in steady-state exponential growth, the growth rate $k$ must correspond to the solution to the equation:
\begin{equation}
1 =   2\tilde{p}(k), \label{eqgrowthlap}
\end{equation}
an equation that is well known \cite{Powell1956}. 

Let $N_{\le j}$ be the total number of cells in states $i = 1...j$.  The dynamics of this quantity has a simple form due to the telescoping form of the dynamics equations  (Eqs.~\ref{eqn:dynamics}-\ref{eqnconv}) where the number entering the $i$th state exactly cancel the number leaving the $i-1$th state:
\begin{eqnarray}
\tilde{N}_{\le j} &=& \tilde{N}_1^+ - \tilde{N}_{j+1}^+, \\
&=& (1-\tilde{p}_{j}) \tilde{N}^+_{1}.
\end{eqnarray}
To determine the overall normalization, we can sum up the cells in all states and set that sum equal to the total number of cells $N(t)$:
\begin{equation}
\tilde{N} = \tilde{N}_{\le m} = \textstyle \frac{1}{2} \tilde{N}^+_{1}.
\end{equation}
From $\tilde{N}_{\le j} $, we can compute the number in individual states:
\begin{eqnarray}
\tilde{N}_{j} &=& \tilde{N}_{\le j}-\tilde{N}_{\le j-1},\\
&=& (\tilde{p}_{j-1}-\tilde{p}_{j})  \tilde{N}^+_{1}.
\end{eqnarray}
In the long time limit, the fastest growing mode dominates the solution and therefore:
\begin{eqnarray}
N(t) &=& N(0)\ e^{-kt},\\
{N}_{j}(t) &=&  2[\tilde{p}_{j-1}(k)-\tilde{p}_{j}(k)] N(t),\\
{N}_{\le j}(t) &=&  2[1-\tilde{p}_{j}(k)] N(t),\\
N^+_{j}(t) &=& 2\tilde{p}_{j-1} N(t),\label{eq2plus}
\end{eqnarray}
which also appear in the main text.

\subsection{The exponential mean of a very narrow distribution}

\label{expmeannarrow}
To obtain Eq.~\ref{eqn:expan}, we begin with the exponential mean:
\begin{equation}
    \bar{t}(k)=-\textstyle{\frac{1}{k}}\ln\mathbb{E}_t\qty[\exp(-kt)],
\end{equation}
which is obtained by using the function ${g(t)=\exp(-kt)}$ in the generalized expectation equation (Eq.~\ref{eq:generalexpectation}). We then use a series expansion of the exponential term, ${e^x=1+x+\frac{1}{2}x^2+...}$, which yields:
\begin{eqnarray}
\bar{t}(k)&=&-\textstyle\frac{1}{k}\ln\qty(\mathbb{E}_t\qty[1-kt+\frac{1}{2}k^2t^2]+...),\\
&=&-\textstyle\frac{1}{k}\ln\qty(1-k\mathbb{E}_tt+\frac{1}{2}k^2\mathbb{E}_t[t^2]+...).
\end{eqnarray}
We then use the series expansion ${\ln(1+x)=x-\frac{1}{2}x^2}$, keeping only second order terms, since the distribution is very narrow:
\begin{align}
\bar{t}(k)&=-\textstyle\frac{1}{k} \big(-k\mathbb{E}_tt+\frac{1}{2}k^2\mathbb{E}_t[t^2]\notag\\
&\qquad\qquad-\textstyle\frac{1}{2}\qty[-k\mathbb{E}_tt+\cancel{\frac{1}{2}k^2\mathbb{E}_t[t^2]}]^2+...\big)\\
&=-\textstyle\frac{1}{k}\qty(-k\mathbb{E}_tt+\frac{1}{2}k^2\qty[\mathbb{E}_t[t^2]-\qty(\mathbb{E}_tt)^2]+...)
\end{align}
Using the definition of the variance, ${\sigma_t^2=\mathbb{E}_t[t^2]-\qty(\mathbb{E}_tt)^2}$, we obtain Eq.~\ref{eqn:expan}:
\begin{equation}
\bar{t}(k)=\mathbb{E}_tt-\textstyle\frac{1}{2}k\sigma_t^2+...
\end{equation}

\subsection{Derivation of the consistency condition in the stochastic model}

\label{stoch_consistcond}
In the deterministic model, Eq.~\ref{eqnidealconst} is a consistency condition that describes the na\"ive expectation that the duration of the cell cycle is equal to the doubling time of the population:
\begin{equation}
    T_\tau=T.
\end{equation}
In the stochastic model, the consistency condition in terms of the Laplace transform is given by Eq.~\ref{eqn:consistgm}:
\begin{equation}
    1=2\tilde{p}(k).
\end{equation}
However, the mathematical equivalence is opaque for the moment. To make the equivalence clear, we use the exponential mean Eq.~\ref{expmeanlife}:
\begin{equation}
\overline{\tau}_j(k) \equiv -\textstyle \frac{1}{k} \ln \tilde{p}_{j}(k),
\end{equation}
along with the relation between the PDF of lifetimes of individual state $j$ and the PDF of times taken to transition from birth through state $j$:
\begin{equation}
\tilde p_{j} = \prod_{i=1}^{j} \tilde p_{\delta i},
\end{equation}
which is Eq.~\ref{eqnconvthr} in the main text. Combining these two equations and letting $j$ be the final state $m$, we obtain the exponential mean of the stochastic cell cycle duration:
\begin{eqnarray}
    \overline{T}_\tau&=&-{\textstyle\frac{1}{k}}\ln\qty(\prod_{i=1}^m\tilde{p}_{\delta_i}),\\
    &=&\sum_{i=1}^m-{\textstyle\frac{1}{k}}\ln\tilde{p}_{\delta_i}=\sum_{i=1}^m\overline{\tau}_{\delta_i}.
\end{eqnarray}
If we use the consistency condition Eq.~\ref{eqn:consistgm}, we can also write:
\begin{eqnarray}
    \overline{T}_\tau&=&-{\textstyle\frac{1}{k}}\ln\tilde{p}(k),\\
    &=&k^{-1}\ln 2=T,
\end{eqnarray}
where the second equality came from the definition of the doubling time (Eq.~\ref{defdoub} in the main text). We thus recover Eq.~\ref{eqsumoft} from the main text:
\begin{equation}
\overline{T}_\tau \equiv \sum_{i=1}^m\overline{\tau}_{\delta i} = T,
\end{equation}
which corresponds to the consistency condition in the deterministic model, Eq.~\ref{eqnidealconst}.

\subsection{A generalization of the stochastic model}
\label{genmodelsec}
Like the deterministic model before it, it seems almost certain that the phenomenology of the stochastic model is more general than some of the assumptions made to motivate and derive it. In particular, the qualitative mechanism that makes the exponential-mean of state lifetime  the determinative statistic would seem to depend only on the exponential enrichment of young cells in an exponential culture and not on the details of the sequential state structure of the stochastic model. We therefore offer a slightly more general derivation below.

In the generalized model, assume only that state or object $j$ is created with wait time distribution $p'$ relative to the birth of a new cell and assume steady-state growth at rate $k$. Under these assumptions, $\tilde p'$ replaces $\tilde p_{j-1}$ in Eqs.~\ref{eq1plus} and \ref{eq2plus}, even if $k$ is not determined by Eq.~\ref{eqgrowthlap} due to memory effects. Therefore, most of our results generalize in this new model if the suitable PDFs for the wait times replace the $p_j$'s.

\end{document}